# Martian north polar cap summer water cycle

Adrian J. Brown[*1], Wendy M. Calvin[2,], Patricio Becerra[3], Shane Byrne[3]

[1] *SETI Institute, 189 Bernardo Ave, Mountain View, CA 94043, USA*

[2] *Geological Sciences, University of Nevada, Reno, NV, 89557, USA*

[3]*Lunar and Planetary Laboratory, University of Arizona, Tucson, AZ, 85721, USA*

Abstract. A key outstanding question in Martian science is "are the polar caps gaining or losing mass and what are the implications for past, current and future climate?" To address this question, we use observations from the Compact Reconnaissance Imaging Spectrometer for Mars (CRISM) of the north polar cap during late summer for multiple Martian years, to monitor the summertime water cycle in order to place quantitative limits on the amount of water ice deposited and sublimed in late summer.

We establish here for the first time the summer cycle of water ice absorption band signatures on the north polar cap. We show that in a key region in the interior of the north polar cap, the absorption band depths grow until $L_s$=120, when they begin to shrink, until they are obscured at the end of summer by the north polar hood. This behavior is transferable over the entire north polar cap,

___________
[*] corresponding author, email: abrown@seti.org

where in late summer regions 'flip' from being net sublimating into net condensation mode. This transition or 'mode flip' happens earlier for regions closer to the pole, and later for regions close to the periphery of the cap.

The observations and calculations presented herein estimate that on average a water ice layer ~70 microns thick is deposited during the $L_s$=135-164 period. This is far larger than the results of deposition on the south pole during summer, where an average layer 0.6-6 microns deep has been estimated by Brown et al. (2014).

Summary. We use CRISM mapping observations of the Martian north polar cap to quantify the deposition of water ice in summer. We have analyzed changes in albedo and water absorption band maps of the entire north polar region as a function of space and time. We identify 'net deposition' and 'net condensation' regions and periods and present a unified model of how these regions change over the course of an entire Martian boreal summer.

Key Point 1. We have identified regions and periods of 'net deposition' and 'net sublimation' on the north cap

Key Point 2. Regions of the cap undergo a 'mode flip' from sublimation to deposition mode

Key Point 3. We estimate water ice layer ~70 microns thick deposited in the $L_s$=135-164 period

Corresponding author:

Adrian Brown

SETI Institute

189 N. Bernardo Ave, Mountain View, CA 94043

ph. 650 810 0223

fax. 650 968 5830

email. [abrown@seti.org](mailto:abrown@seti.org)

Short running title: "Summer $H_2O$ cycle on Martian north polar cap"

## 1. Introduction

The Martian water cycle is crucial to the understanding of the dynamics of the atmosphere, surface and sub-surface of the planet (Clifford, 1993). Since the Viking mission, the north polar cap has been understood as the key source and sink of modern day Martian water ice and vapor (Farmer et al., 1976; Kieffer et al., 1976). The sublimation and deposition of water ice and dust have also created an important record of layering in the North Polar Layered Deposits (NPLD) that could unlock the history of the Martian climate  (Fishbaugh and Head, 2005). Recent SHARAD mapping of the subsurface structure of the NPLD (Putzig et al., 2009) has revealed a rich history of deposition of dust and water ice.

### 1.1 General approach of this study

The approach of this study is to use CRISM observations of the North Polar Residual Cap (NPRC) during summer to quantify the deposition and sublimation of water ice under the current Martian climate. We use CRISM spectra and a set of mosaic maps created specifically for this project to track spatial and temporal changes in the depth of water ice absorption bands. Our dataset spans multiple Martian years, and we use radiative transfer models to infer a snowpack composition that is consistent with the spectra. In this way, we can address the key Martian climate question of how much water ice is deposited on the north polar cap during summertime.

In this study, we map a key spectral parameter called the $H_2O$ ice index, which is sensitive to the 1.5μm water ice absorption band. As discussed in more detail below, the ice index indicates when water ice is present, and the strength of this index is high when large grained ice is present.

*1.1.1 Detection of water ice deposition events*

Fine grained ice has a lower $H_2O$ ice index than large grained ice. We can therefore use relative changes in the $H_2O$ ice index to track deposition events on the north polar cap. This strategy is optimal for large scale surveys of the water cycle across the north polar cap. However, it does have the drawback that it is insensitive to the type of water ice deposition. Both snowfall (precipitation) and direct deposition of water ice from the atmosphere produce a similar effect on the $H_2O$ ice index. In addition, due to the overflight frequency of MRO, our approach is insensitive to the time of day of the deposition events. We discuss the advantages and limitations of our approach in more detail below.

*1.2 Previous Observations of the NPRC water cycle*

Observations of the water cycle on the north polar cap have a long history. Trying to interpret Viking Infra Red Thermal Mapper (IRTM) observations, Kieffer (1987) first suggested the possibility of water ice being deposited onto the north polar cap to explain summer brightening of the albedo of the cap. Kieffer (1990) subsequently used a thermal metamorphism model and albedo calculations of ice-dust mixtures to infer that the north polar cap must be composed of either 1.) "old, coarse and clean" or 2.) "young, fine and dirty" ice. The available IRTM data could not discriminate between these two scenarios. As discussed in Section 5 below, the present study has revealed new subtleties that reach beyond these two scenarios of Kieffer (1990). CRISM mapping has identified regions and periods where the water ice of the exposed north polar cap undergoes 'net deposition' and 'net sublimation' modes across the entire cap, as fine grained ice is deposited on top of relatively coarse grained ice.

*1.2.1. Viking and Mariner 9 Observations of the north polar cap*

Using Mariner 9 and Viking imagery, Bass et al. (2000) determined that the albedo of the polar cap brightened during late summer, after darkening in early summer. They performed Mie sphere-based radiative transfer (RT) modeling to reproduce the observed albedos. They could not determine the size of the $H_2O$ ice grains or entrained dust grain sizes on top of the cap because the Viking and Mariner cameras provided only broadband visible albedo measurements.

Bass et al. (2000) suggested that in order to match their RT model to their broadband albedo observations, fresh layers of water ice must be deposited during late summer - with deposition depths that total at least 14 microns (for small grains) or 25 microns (for large grains) per season. This compares relatively well with the earlier findings of Hart and Jakosky (1986) who found that at the Viking 2 lander site, a decrease of 5% in reflectance corresponded to 10 microns of water ice deposited during one Martian summer.

Bass et al. (2000) critically examined Kieffer's two scenarios of 'old, coarse and clean' vs. 'young, fine and dirty' for the north polar snowpack. They preferred the explanation of deposition of new ice rather than creation of suncaps or cracks (Bass et al. p. 390) because cracking would require the preservation of older ice with lower albedo, which they considered less likely. Bass et al. also established that the albedo brightening was not due to cloud activity by establishing that there is no increase in albedo from dark areas adjacent to the bright cap.

Bass and Paige (2000) used Viking IR thermal mapper (IRTM) and Mars Atmospheric Water Detector (MAWD) measurements to determine the peak amount of water vapor above the north polar cap. They found that the cap visible albedo was lowest between $L_s$=93-103 and significant amounts of water vapor began to be released around $L_s$=103. This was accompanied by an increase in cap albedo. Observed surface temperatures of the cap were too warm for re-deposition of $CO_2$ ice. They presented a model (their Fig. 11) of the NPRC ice cap as a cold trap for water vapor. Under this model, sublimation of water ice during spring and summer builds up water vapor in the atmosphere until $L_s$=103, after which time the water vapor is deposited as fine grained $H_2O$ ice accumulated on top of the cap, as temperatures cool in late summer. We will test and extend this depositional model hypothesis in Section 5 of this paper.

### 1.2.2 Mars Global Surveyor Observations of the North polar cap

Kieffer and Titus (2001) utilized Thermal Emission Spectrometer (TES) data to map the retreat of the north polar seasonal cap using surface temperature measurements. The main focus of their study was the dynamics of $CO_2$ ice, however they did observe increases in albedo which they attributed to deposition of fine grained water ice. In addition, Kieffer and Titus tracked a bright but warmer (i.e. non-$CO_2$ ice) annulus that tracked the retreating $CO_2$ cap, and suggested it was a water ice annulus.

Pankine et al. (2010) used TES measurements to determine the amount of water vapor over the north pole during the northern spring and summer of Mars Year (or MY, for details see (Clancy et al., 2000)) 24 to 27. Pankine et al. found that the amount of water vapor peaks in summer over the north polar cap between $L_s$=105-115 at 80-90 precipitable microns (pr-µm) over the 85-87° latitude range. These results showed a good match for the Bass et al. observations but also displayed a wider range of latitudinal-dependent timings for peak values of water vapor.

### 1.2.3 Mars Express Observations of the North Polar cap albedo

Langevin et al. (2005) used OMEGA near infrared (NIR) observations to document the north polar cap albedo during early summer ($L_s$=93-127). They found that the NIR albedo decreases, and attributed this as due to the growth of water ice grains. Langevin et al. interpreted this increase in grain size as due to fine grained ice being stripped from the NPRC and thereby exposing older, large grained ice to the surface. It is also likely that thermal metamorphism (Eluszkiewicz, 1993; Ossipian and Brown, 2014) plays a role in the increase of grain sizes during early Martian summer. Following the Langevin et al. finding that the NPRC is currently losing mass, Byrne (2009) suggested that "due to its thin nature this situation cannot persist for long". The residual ice cap on top of the NPLD is thought to be approximately one meter thick (Byrne, 2009). It is thus important to establish the full summer sublimation/condensation cycle to understand the stability of the top layers of the NPLD.

### 1.3 Previous Numerical Modeling of the NPRC water cycle

Numerous computational models have been used to investigate questions of water ice transport and deposition of volatiles on the Martian polar caps, with varying degrees of physical fidelity. For example, on the less complex side, Haberle and Jakosky (1990) used a 2D transport model to assess the ability of the north polar cap to supply water to the atmosphere. Simple one dimensional energy balance models have also been used to investigate the springtime eruption of geysers in the south polar region (Kieffer, 2007; Pilorget et al., 2011).

On the more complex side, Global Circulation Models (GCMs) have also been used to address polar energy balance questions. For example, Pollack and Haberle (1990) used a 3D GCM to investigate the effect of dust on the condensation of $CO_2$ ice in the polar regions. Houben et al. (1997) used a simplified GCM and discovered that cold trapping on the retreating seasonal cap was an efficient process for moving water ice to the NPRC (see also Titus et al. (2008)).

*1.3.1 Mesoscale modeling of the NPRC water cycle*

Standard GCMs generally have difficulty mapping the polar regions. There are two key reasons for this. The first difficulty is the so-called "pole problem" that is due to the fact that standard GCMs use rectangular coordinates which have a singularity at the pole where the gridpoints converge. In these models it is necessary to use polar filters to smooth out regions close to the pole (Lian et al., 2012). The second difficulty is spatial resolution – a typical global Mars GCM is run at 5º of latitude and 6º of longitude, and this is simply insufficient to capture the complex topography, cloud patterns and wind vectors in the north polar region (Tyler Jr. and Barnes, 2014). As a result, mesoscale modeling is required in order to better understand the dynamics of the Martian polar regions. Mesoscale models achieve higher resolution over a certain region of the planet, making them very CPU-intensive computer programs that require human intervention and judgement. For example, their boundary conditions are required to be sourced from 'suitable' global scale GCM runs.

Tyler and Barnes (2005) reported the results of first mesoscale modeling in the Martian polar regions. They conducted three mesoscale simulations of the entire north polar cap at different periods in the Martian summer: $L_s$ = 120, 135 and 150. In this model, the atmosphere was dry and therefore sublimation and condensation rates of water ice were not simulated. Subsequently, Tyler and Barnes (2014) published the results of a "wet" mesoscale model which included clouds (which were not, however, radiatively active) and a water ice condensation/sublimation scheme. This model simulated a period of 20 sols (Martian days) after $L_s$=120, and had a highest resolution grid of 15km on the surface. Their key findings (for the purposes of this paper) were as follows:

* they produced a 20 sol diurnal mean sublimation map of the polar cap at 15km resolution, showing highest sublimation rates around the periphery of the cap, peaking at 50 microns around the cap edge (their Figure 12a),

\* they presented a condensation map of the water ice deposition by snowfall and direct deposition, showing deposition mostly on the interior of the cap, peaking at 5 microns per day (their Figure 12b).

\* they calculated the net condensation on the cap to be 6% of the net sublimation of the cap at this time, reflecting the relatively warm mid summer period at $L_s$=120.

The key findings of this recent numerical study allow us to put the results of the current study in better context, and also to determine the locations and periods where observations and models currently agree and disagree.

## 2. Methods

### 2.1 Instruments

We have used observations acquired by two separate instruments to conduct our study of the north polar cap. The Compact Reconnaissance Imaging Spectrometer for Mars (CRISM) is a visible to near-infrared spectrometer on the MRO spacecraft that is sensitive to near infrared (NIR) light from ~0.39 to ~3.9μm and is operated by the Applied Physics Laboratory at Johns Hopkins University (Murchie et al., 2007). The MARs Color Imager (MARCI) is a super wide angle, fish eye lens camera with 1024 pixels-wide CCD, also on MRO, that is operated by Malin Space Science Systems (Malin et al., 2001).

In the CRISM mapping mode, 10x binning is employed in the cross-track direction, consequently the mapping images used herein are 60 pixels across, covering approximately 10.8km on the surface with a ground resolution of ~182m per pixel. The length of each swath is controlled by exposure time and is variable depending on commands sent to MRO.

We produced 1000x1000 pixel mosaics of all the CRISM mapping data available for each two week period (equivalent to one period in the MRO planning cycle). We performed a "cos(i)" correction to account for variable illumination angles for all CRISM data. Each mosaic is presented in polar stereographic projection and stretches down to 72°N at the cardinal points. Figure 1 shows the area covered by each mosaic as a MARCI image of the north polar cap.

*2.2 H$_2$O ice index*

Surface water ice can be mapped by applying spectral band fitting techniques (Brown, 2006; Brown et al., 2008b) on the near infrared mosaics and exploiting the strong 1.25µm and 1.5µm water ice absorption bands (Bibring et al., 2004; Langevin et al., 2005; Brown et al., 2012).

Throughout this paper, we will refer to a spectral parameter we call the "H$_2$O ice index" that was first suggested by Langevin et al. (2007) and later adjusted for use with CRISM by Brown et al. (2010). The formula for the index is:

$$H_2O index = 1 - \frac{R(1.5)}{R(1.394)^{0.7} R(1.75)^{0.3}}$$
(1)

where $R(\lambda)$ indicates the reflectance at the wavelength $\lambda$ in μm.

### 2.2.1 Physical basis of the H$_2$O ice index

The H$_2$O ice index is based on the depth of the water ice 1.5μm absorption band. When water ice is present, the 1.5μm absorption band appears and thus the *R(1.5)* factor in the second term decreases. Since this factor is subtracted from 1, the overall index therefore increases. As water ice grains grow, the *R(1.5)* factor continues to decrease and the index continues to grow. The index is therefore higher when large grained water ice is present (Warren, 1982; Brown et al., 2008a). When deposition of fine grained ice occurs, the *R(1.5)* factor increases (and the H$_2$O ice index decreases), because finer grained ice scatters light back to the observer more readily (Warren, 1982) and this in turn decreases the depth of the 1.5μm H$_2$O ice absorption band. Thus over the course of a Martian summer, when fine grained water ice is deposited onto the north polar cap, we should recognize a trend for increasing H$_2$O ice index. We will show in the results section that we can indeed identify this trend and use it to detect and map deposition events.

## 2.2.2 Error Analysis of the $H_2O$ ice index

For this study, we have made no effort to remove atmospheric effects from the CRISM data. We will now examine the possible errors in the derived $H_2O$ ice index that could result from this processing simplification. Our desired goal is to determine the chance that we have made an error when we claim to have detected a change of 0.1 in the $H_2O$ index in our subsequent analysis.

### 2.2.2.1 Effects of Instrumental Noise on $H_2O$ Ice Index

In the near infrared spectral region around 1.5 µm, the uncertainty introduced by instrumental noise is minimal, since CRISM signal to noise (SNR) at 1.5 µm is ~400 (Murchie et al., 2007). The noise profile of the CRISM instrument has changed over time, however, for this study, we use observations from the 2006-2010 period, when the instrument was still in an extremely healthy state. Therefore we assume the SNR=400 over the region of the $H_2O$ index. To claim detection of the desired $H_2O$ index change of 0.1 would then lead to errors (1/400)/0.1 = 2.5% of the time.

### 2.2.2.2 Effect of haze and clouds on $H_2O$ Ice Index

We wish to quantitatively examine the rise and decline of the $H_2O$ index, therefore the effect of fine grained water ice clouds in the Martian north polar region on this parameter is critical to understand. We carried out a study of CRISM observations of the Phoenix site (described fully in the Appendix) and arrived at an error estimate for the atmospheric effects on the $H_2O$ index of ± 0.01. This roughly equates to a potential 10% error due to atmospheric effects, since the significant findings discussed in the Observations (Section 3) below rely on changes of around 0.1 in the $H_2O$ index.

For a typical Martian north polar spectrum, the presence of a saturated 1.5µm absorption band is indicative of ice grains greater than 100 microns in diameter on the surface of the planet (Brown et al., 2008a, 2012). Water ice grains of this size would not have a long residence time in the thin Martian atmosphere, therefore they cannot be confused with fine grained water ice clouds, such as those detected above the south polar cap by CRISM in early spring (Brown et al., 2010).

In addition, Clancy and Lee (1991) conducted a study using Viking IRTM emission phase function data and found that brighter (icy, 50-75% albedo) surfaces are less affected by atmospheric interference than dark surfaces (of 10-30% albedo). They found bright surfaces on Mars appear typically only 10% darker that their true albedo when observed by orbiting spacecraft. This agrees well with our assessment of a potential 10% error in the $H_2O$ index due to atmospheric effects.

*2.2.2.3 Effects of dust on the $H_2O$ ice index*

Recent studies have also highlighted the role of atmospheric dust in affecting the apparent albedo of the north polar cap. Cantor et al. (2011) used MARCI visible image data to track dust storms and changes in the albedo of the north polar cap. They showed that traveling spiral dust storms often intersect the cap edge during summer. For example, Figures 18 and 20 of their paper show the process of dust deposition causing a rapid decrease in albedo (especially on the edge of the cap) during storms in late summer. Therefore, dust has the effect of darkening the polar cap, whereas in this study we are interested in the phenomenon of cap brightening.

Calvin et al. (2014) reported on MARCI observations of similar albedo variations over three summers (MY29-31) and noted a decrease in albedo up to $L_s$=95, followed by brightening over large regional areas, especially the Gemini Scopuli region colloquially known as "Shackleton's Grooves". Wide interannual variability was noted in that paper over smaller sections of the NPRC such as Abalos Mensae, Olymipia Cavi and Olympia Undae, and the lower elevation fringes of Gemini Lingula, but that study did not explore the albedo increases on the higher elevation NPRC residual ice discussed in this paper.

Summarizing these results we can say that the effects of dust on the $H_2O$ index are difficult to quantify. However we can state that because our major results rely on observations of increasing albedo (which is the opposite effect expected for dust deposition) we believe our findings will remain robust despite the presence of mobile dust on the NPRC.

*2.2.2.4 Total error estimate for H₂O ice index*

We therefore have been able to estimate the affects of instrumental (±2.5%) and atmospheric (±10%) effects. Assuming these errors are statistically independent, and using the relation *total error = sqrt(error$_{instrumental}$² + error$_{atmospheric}$²)* the instrument error gives only an additional 0.31% to the atmospheric error. Rounding to one significant digit, we use ±10% for the relevant error estimate when quoting values for the $H_2O$ ice index later in the text.

*2.3 Timing of North Polar Hood Obscuration*

In the north polar region, we have known since the Viking mission of thin spiral vortex clouds that appear to form near the cap edge and move south toward the equator during summer (French and Gierasch, 1979; Kahn, 1984). Multi-year observations of the north polar cap during summer (Cantor et al., 2002; Tamppari et al., 2007; Wang and Ingersoll, 2002) have shown spatial and temporal repeatability in several cloud forms which has lead to the identification of a "storm zone" that is active in late summer between Alba Patera and the north polar cap.

Our observations of MARCI mosaic images for MY28 and 29 over the north polar cap show that the early summer period ($L_s$=100-120) is almost completely clear of clouds. After $L_s$=120, the aforementioned patchy baroclinic (vortex) clouds appear. By $L_s$=140, there are multiple cloud systems around the cap, though apparently not over the NPRC itself. By $L_s$=150, a thin haze of cloud envelops the entire pole out to around 60°N. Around $L_s$=160-165, the haze rapidly changes into the thick cover known as the north polar hood. Benson et al. (2011) used the Mars Climate Sounder infrared instrument on MRO to show that the transition to the thick hood occurs at $L_s$=165 in the thermal infrared (see especially their Figure 3b and c).

We have carried out an analysis (described fully in the Appendix) to estimate when the north polar hood has become too optically thick for reliable quantification of ice deposition. Our best estimate is that the polar hood becomes obscuring in the near infrared spectral region at $L_s$=165. This is in good agreement with the results of Benson et al., despite the fact that MCS is sensitive to longer wavelengths (thermal infrared) than CRISM (near infrared). Note that the visible observations (such as those measured by MARCI) are affected more rapidly by the polar hood – the visible albedo is markedly obscured by around $L_s$=150.

## 3. Observations

### 3.1 Spatial Analysis - Grain Size and Albedo Maps

Pole-wide maps of the $H_2O$ ice index and albedo are presented in Figure 2 and 3. High values are shown in red and low values are shown in blue. These show that the $H_2O$ index decreases over large regions of the central cap in the $L_s$=132-167 period and outlines the spatial extent of this effect. It is readily apparent from the red color that changes to green in the later mosaics that the $H_2O$ index decreases across the entire cap during this time. On the periphery of the cap, a blue region of low $H_2O$ index persists throughout late summer.

### 3.1.1 Depositional patterns

As can be seen from the CRISM mosaics, the $H_2O$ index decrease is confined to the boundary of the north polar cap, and does not extend beyond the cap edge. This argues against an atmospheric origin for this process, since CRISM has previously been shown to be capable of detecting thin Martian water ice clouds extending beyond the edge of the cap in the southern polar region (Brown et al., 2010).

### 3.2 Temporal Analysis – Spectral time series from 'Point B'

Unfortunately our CRISM mosaic coverage does not extend over the complete summer period, however to bridge this gap we have examined individual CRISM images that have been obtained repeatedly on some important locations on the cap. One such point is called 'Point B'. Figure 4 shows a time series of individual spectra (no averaging has been done) from 45ºE, 85ºN (point 550, 550 on our 1000x1000 pixel mosaics). This is as close as we could get to a point examined by Langevin et al. (2005) which they termed 'Point B' at 42.5ºE, 85.2ºN. This is very close to the point called 'McMurdo' in the Calvin and Titus (2008) study, and was also examined in Figure 7 of the Byrne et al. (2008) study. This location was also called the 'Cool and Bright Anomaly' (CABA) region by Kieffer and Titus (2001). The two points, 'Point B' and 'McMurdo' are both on the spur connecting Planum Boreum (the main part of the cap) to Gemini Lingula (the tongue below the main body in Figure 1). The Figure 2-3 mosaics support the assumption that spectra of 'Point B' and 'McMurdo' behave similarly during late summer, therefore we have some justification for only conducting an exhaustive study of the spectral time series from 'Point B' in this study.

### 3.2.1 Timing of NIR albedo increase at Point B

Our best estimate of the timing of the albedo increase at Point B comes from Figure 4, which shows all the available NIR spectra for MY29 at this location. The spectra decrease from $L_s$=90-115, then hold steady until $L_s$=135.25 when the spectrum amplitude begins to increase again. Observations are ended when the polar hood fully obscures the cap.

It is of interest to note the albedo at ~ 1 μm remains steady across the late summer, as has been reported by Calvin and Titus (2008) and Byrne et al. (2008). This is because for a water ice spectrum, the 1 μm region is less sensitive to grain size changes than the NIR (Grenfell et al., 1981, Figure 3) particularly in the presence of contaminants, which serve to control the visible albedo (e.g. on Mars). This is discussed further below (Section 4).

*3.3 Interannual analysis - Variations in $H_2O$ index*

The $H_2O$ ice index tracks the depth of the 1.5 μm absorption band, so as the NIR albedo decreases in early summer, the $H_2O$ ice index increases. In late summer, the albedo increases and the $H_2O$ ice index decreases. In order to establish that this process is a multi-year cycle, we have extracted individual spectra from regions close to Point B and have plotted the $H_2O$ ice index for these spectra in Figure 5. Observing the $H_2O$ ice index curves from MY28, 29 and 30, one can see that the $H_2O$ index at Point B behaves in a similar manner (i.e. it increases until $L_s$=120, then decreases) through summer across multiple Martian years.

**4. Radiative Transfer Modeling**

We have carried out radiative transfer modeling using a model of the Martian snowpack based on DISORT (Stamnes et al., 1988), which is a layered, multiple scattering model that allows us to approximately quantify the composition of the observed snow pack using CRISM spectra. The fitted spectra were all taken from Point B in MY28 in mid to late summer and are shown in Figures 6a and b.

The aim of our modeling is to constrain the question – what amount of deposition of fine grained ice will reproduce the spectral changes we see between $L_s$=135.25 (mid summer, when the NIR spectra at Point B begin to increase) and $L_s$=163.85 (late summer)? Since we have the $H_2O$ ice index for both these spectra, we will use our model to estimate how much water ice deposition corresponds to the observed change in the $H_2O$ ice index.

*4.1 Near infrared and visible albedo decoupling of water ice*

Pure water ice is extremely transparent in the visible and extremely absorbing in the near infrared. This leads to a decoupling of visible and near infrared spectral responses. When contaminants are present in ice (as in a dusty snowpack) the *visible* albedo is greatly diminished. However, in the *near infrared*, where water ice is strongly absorbing, the scattering properties of the contaminant can only have a minimal effect on the NIR albedo of the mixture. Thus, the controlling factor in the NIR is the grain size of the highly absorbing water ice. This decoupling effect has been known for some time (Warren and Wiscombe, 1980) and has been observed in the Antarctic (where the water ice is much cleaner than the Martian north pole and visible albedos are higher, usually > 0.9) (Grenfell et al., 1994). This decoupling is the key to why CRISM NIR observations can be confidently used to determine ice grain sizes – by fitting our model to the NIR spectrum, any variations that we see are largely due to grain size changes and not the product of contamination by dust.

*4.2 Snowpack Model*

The task of modeling the infrared albedo of an ice pack is challenging (Warren, 2013; Becerra et al., 2015; Kaufmann and Hagermann, 2015), therefore we simplified our model as much as possible and set our goals as conservatively as possible. Extensive terrestrial snowpack radiative transfer studies exist using DISORT to represent a coupled snowpack-atmosphere system (Jin et al., 1994; Hamre et al., 2004; Stamnes et al., 2011). In our implementation, because the Martian atmosphere is so thin, we conducted no atmospheric removal, as discussed earlier.

Following previous terrestrial studies (Grenfell, 1991; Stamnes et al., 2011), to generate our snowpack model, we use the simplifying assumption that each Martian water ice grain scatters like a sphere. We are then deriving an "equivalent radius" for that sphere in our model. The reason why this approach is successful in the terrestrial context was described in this way by Craig Bohren, (also quoted in (Grenfell et al., 1994)):

"The orientationally averaged extinction cross section of a convex particle that is large compared with the wavelength is one-half it's surface area. The absorption cross section of a large, nearly transparent particle is proportional to its volume almost independent of its shape. The closer the real part of the particle's refractive index is to 1, the more irrelevant the particle shape. The asymmetry parameter of large particles is dominated by near-forward scattering, which does not depend greatly on particle shape."

In addition, as is the common practice when modeling terrestrial snowpacks (Hamre et al., 2004; Stamnes et al., 2011), we used a Henyey-Greenstein (HG) phase function within DISORT, and fed the HG phase function with an asymmetry parameter derived from Mie calculations. Unlike Stamnes et al. (2011), we used actual Mie calculations rather than geometrical optics derived parameterizations.

Our modeling goal is to quantify the amount of fine grained water ice deposited between two points in time. To that end, we created a three-layered model using 1.) fine grained ice, 2.) coarse grained ice and 3.) dust. We used the optical constants of water ice (Warren, 1984) and palagonite (Roush et al., 1991) as a Martian dust analog. The parameters of our DISORT snowpack are given in Table 1.

### 4.2.1 $L_s$=135.25 spectrum

In the first run, we attempted to match the MY28 $L_s$=135.25 spectrum from Figure 6a. For this run only, we omitted the fine grained ice layer, and fitted the spectrum with only coarse gained water ice covered by dust. We carried out a constrained iterative fit, as described in (Brown et al., 2014) utilizing the Nelder and Mead simplex algorithm (Nelder and Mead, 1965), and using the chi-squared $\chi^2$ metric to constrain our fit.

We used the Nelder and Mead simplex algorithm to find a good fit to the spectrum by varying the diameters and relative quantities of the coarse grained ice and dust components. We found a good fit to the data was achieved when the coarse grain diameters were ~1100 microns with an optical depth of 2.36. The dust was best fit by an effective grain diameter of ~143 microns with an optical depth of 1.56. These results are summarized in Table 1.

The best fit of the effective grain diameter of the dust particles is two orders of magnitude greater than the diameter of typical atmospheric-borne Martian aerosols (~1-5 microns, Wolff et al. 2009). This suggests that the grains responsible for the darkening of the north polar snowpack at Point B must have undergone coagulation, or may in fact be sand sourced from the dunes surrounding the cap. It should be borne in mind that we are quoting an average value and we have made no attempt to constrain the true size distribution, which would involve significant assumptions and complications that are not warranted for the purposes of this study.

*4.2.2 $L_s$=163.85 spectrum*

In our second run, we tried to model the $L_s$=163.85 spectrum using a ceteris paribus assumption (Reutlinger et al., 2011). Our "ceteris paribus" assumption attempts to change just one variable, while leaving everything else the same. We therefore attempted to keep the coarse grained ice and dust diameters roughly the same as in the first fit, however some changes were necessary to achieve a good fit.

In our second run, we introduced a fine grained layer on top of the dust and coarse snow. We fixed our fine grained ice component to an effective diameter of ~50 microns because this is the most recent estimate of typical average water ice grain sizes within Martian north polar clouds (Lemmon, 2014) and may thus be a good size for freshly deposited snow particles.

When we carried out the fit to the $L_s$=163.85 spectrum, we found that the best fit was achieved when we used a fine grained (diameter of 53 microns) water ice layer with an optical depth of 1.53. The results of the fits are also summarized in Table 1.

| Component | Reference | $L_s$=135.25 $H_2O$ index=0.692 ($\chi^2$ fit=1.567) | | $L_s$=163.85 $H_2O$ index=0.495 ($\chi^2$ fit=0.769) | |
|---|---|---|---|---|---|
| | | Effective Grain diameter (microns) | Optical depth | Effective Grain diameter (microns) | Optical depth |
| Coarse $H_2O$ ice | Warren (1984) | 1101.2 | 2.36 | 1813.3 | 0.80 |
| Fine $H_2O$ ice | Warren (1984) | 0.00 | 0.00 | 53.3 | 1.53 |
| Palagonite (soil) | Roush et al. (1991) | 142.6 | 1.56 | 150.1 | 2.00 |

| DISORT Snowpack Model Parameters | |
|---|---|
| Number of layers | 3 |
| Number of streams | 32 |
| Type of Phase Function | Henyey-Greenstein |
| Number of Legendre Coefficients in phase function expansion | 33 |
| Number of computational polar angles | 4 |
| Bottom surface | Lambertian, alb=1 |

Table 1 – Details of the best fit parameters for the DISORT snowpack to the CRISM spectra in Figure 6. The fit was applied over the spectral range from 1.02-2.5 microns. In this range, using wavelengths from the CRISM MSP observations, there are 42 CRISM channels. The $H_2O$ index for each spectrum and the $\chi^2$ fit for each is also provided.

*4.3 Quantification of deposition*

In order to estimate how much water ice has been deposited on the north polar cap between $L_s$=135 and $L_s$=164 it is necessary to make some simplifying assumptions. We adopt a simple model by which we can convert the optical depth in the snowpack to the height of a layer of water ice using equation 11 of Stamnes et al. (2011):

$$h = \frac{\tau}{\pi r^2 N (Q_a + Q_b)} \tag{2}$$

where $h$ is the height of the snowpack, $r$ is the effective radius of the particles (here 26.7 microns), $\tau$ is the derived optical depth ($\tau = 1.53$ for the fine grained ice layer), and $Q_a$ and $Q_b$ are the absorption and scattering coefficients of water ice. We take $Q_a$ and $Q_b$ to be their values at around 1.5 μm, $Q_a = 1$ and $Q_b = 1.5$. $N = (1/(4/3)\pi r^3)\rho_s/\rho_i$ is the total number of particles per unit volume. $\rho_s$ is the density of snow and $\rho_i = 0.917$ g/m$^3$ is the density of water ice. Freshly fallen terrestrial snow typically has $\rho_s = 0.1$ g/m$^3$ and therefore $N = 4516.6$. Putting these values into equation 2, we get a height of 5.6 cm of freshly fallen snow at Point B in late summer. If we instead assume that the water ice is directly condensed onto a relatively dense compacted snowpack (in this case typically $\rho_s = 0.8$ g/m$^3$) we derive a height of 7.05 mm. We will work with the latter estimate going forward, although it should be emphasized that we have no evidence at this time that the freshly deposited snow scenario is incorrect. In situ or experimental measurements of Martian snowpack formation under realistic conditions would be of great benefit to improve this model.

When we take into consideration that the change in $H_2O$ index at Point B over the same time is 0.692-0.495 = 0.197, we are lead to an estimate of 35.7 mm of deposition per unit change in $H_2O$ index.

To convert the thickness of the snowpack change into an estimate of the mass of water deposited (in kg) we use the simplified linear relationship (non linear terms are not considered here), with the aforementioned $H_2O$ ice index to snowpack depth change factor of 35.7mm explicitly included:

$$mass_{deposited}(\text{in } kg) = \alpha \, \Gamma \, (35.7) \Delta \, x \, 10^3 \qquad (3)$$

where α is the area of the cap (estimated as 1.5 million sq. km), Γ is the percentage of the cap covered by the condensing or subliming regions and Δ is the average value of the $H_2O$ index for condensing or subliming regions (Γ and Δ are given in Table 2). We will now apply this estimate to the entire cap using the observed changes in $H_2O$ index over the late summer period.

*4.3.1 Net Deposition and Sublimation Spatial Regions*

Figure 7 shows difference maps displaying the change (or delta) in $H_2O$ index between each mosaic in Figure 2. Because the difference maps only show information in regions where both times had data, fewer data points are available in these mosaics. Therefore, to create these difference maps, an interpolation method was used in order to increase the coverage of each CRISM swath. Two runs of interpolation were executed, each run copied the value of immediately adjacent pixels to any blank pixels. No data was discarded in this process.

The bottom right image (Figure 7d) shows the $H_2O$ index difference across the entire set of late summer mosaics. Assuming that the observed change in $H_2O$ index is due to deposition or sublimation of fine grained water ice, we can capture the cross-cap variability of the seasonal deposition and sublimation of water ice and better quantify the amount of water ice transported around the cap. Here, red indicates a decrease in $H_2O$ index (net deposition region), blue indicates an increase in $H_2O$ index (net sublimation region).

We therefore use Figure 7d, which gives the total $H_2O$ index change across the period of $L_s=132$ to 167 in MY28 to identify 'net deposition' and 'net sublimation' regions. We calculate pixel by pixel the average $H_2O$ index change across this area. We then transform this into an equivalent thickness of water ice deposited using equation (2). Finally, we use equation (3) to transform that thickness into an estimate of the mass of water ice deposited or sublimated per day on the entire cap. The results are summarized in Table 2.

| Region Type | Percent covered of cap ($\Gamma$) | Weighted Avg $H_2O$ index per pixel ($\Delta$) | Change in thickness of snowpack over period | Avg surface water change on cap ($10^9$ kg/day) |
|---|---|---|---|---|
| Net deposition | 68.8 % | 0.117 | 4.2mm | 0.079 ± 10% |
| Net sublimation | 31.5 % | -0.080 | -2.8mm | -0.054 ± 10% |

Table 2 – Details of the net changes in $H_2O$ index from Ls=132-167 during MY28 for regions experiencing net deposition and net sublimation. Regions poleward of 87º are excluded. We have calculated the spatial area of the cap covered by each type, taking into account the polar projection of the data. We have used this area to compute a weighted average $H_2O$ index per pixel and then summed the index values for each pixel. We have used the results of our RT model to convert the $H_2O$ index change into the net water transported in or out of these regions in $10^9$ kg/day. The total cap area (excluding regions poleward of 87ºN) is taken to be 1.5 million sq. km (Brown et al., 2012).

$L_s$=135.25 corresponds to Mars sol 289 and $L_s$=163.85 corresponds to Mars sol 343 of MY 28. Therefore, over a period of 54 Martian days, $5.15 \times 10^9$ kg of water ice was deposited on the polar cap, on average $0.079 \times 10^9$ kg per day. In the same period, $0.054 \times 10^9$ kg were lost from regions of sublimation per day. The error bars on these estimates are ± 10%, as derived in Section 2.2.2.4.

In Figure 7, we can see that the depositional areas are concentrated in the central regions of the cap during late summer. It is also likely that the region CRISM cannot observe (poleward of 87ºN) is a relatively large depositional region. By excluding this region, we are effectively providing a low estimate of the amount of water ice deposited on the polar cap. If we simply assume this region has the same deposition behavior seen at Point B, with roughly 7mm of deposition over 54 days, (or ~140 microns per day) this would add a further $0.039 \times 10^9$ kg of water, to arrive at a total of $0.118 \times 10^9$ kg per day deposited on the cap during late summer.

## 5. Discussion

*5.1 Possible Geophysical Scenarios to Explain Water Ice Observations*

In this study we have established for the first time that strong changes in the $H_2O$ index occur across the north polar cap during late summer. These changes can be used to separate the cap into regions based on the growth or decrease in the $H_2O$ index. We shall now briefly consider alternative scenarios or contributing factors to explain the observed changes in the $H_2O$ index.

*5.1.1 Scenario #1 Physical break up*

One possible explanation for the spectral changes is that the large grains of water ice are being decimated in the course of the summer, causing cracks and asperities to appear, thus increasing the number of scattering centers in the ice (which is equivalent to decreasing the grain size). However, this process requires an established old snowpack and infers no fresh water ice deposition, which runs counter to physical models of the cap water cycle dynamics (Haberle and Jakosky, 1990; Tyler and Barnes, 2014) and so is not our preferred scenario.

*5.1.2 Scenario #2 Removal of covering of coarse grained ice to reveal fine grained ice substrate*

A second alternative is that large grain water ice is being removed, thus exposing fine grained water ice that is stratigraphically below it. Since typical terrestrial snowpacks increase rather than decrease in grain size with depth (Carmagnola et al., 2013), this is not our preferred explanation.

### 5.1.3 Scenario #3 Accumulation of fine grained water ice

The third scenario we discuss is that of accumulation of fine grained water ice on top of large grained water ice during the late summer period. This could take place in the form of direct condensation from water vapor in the atmosphere and/or as precipitating snowflakes. This is consistent with MAWD (Farmer et al., 1977), TES (Smith, 2004) and CRISM (Smith et al., 2013) observations that show a decrease in north polar water vapor after $L_s$=120.

### 5.2 Direct Deposition from water vapor vs. precipitation by snowflakes

Making the assumption that the increases in $H_2O$ ice index we observe are due primarily to accumulation of fine grained ice, we still cannot directly attribute the change in $H_2O$ index to a particular deposition process. Both *precipitation* in the form of water ice snowflakes and direct *condensation* of water ice onto the cap can reproduce the signal that we have detected. We can only state that we are observing a surface process (and not a buried process) because our detection method is only sensitive to the top few centimeters of the snowpack. In addition, we cannot say anything about the timing of the diurnal deposition patterns, we can only quantify the diurnal patterns and sum them to obtain the net seasonal deposition or sublimation amount.

Direct deposition from water vapor is consistent with the Bass et al. (2000) model of water ice subliming on the warm exterior of the cap traveling poleward, and sinking and accumulating on the cold cap interior. The other alternative of ice deposition by snowflake deposition is supported somewhat by the requirement in our RT model to add a fine grained water ice component to fit the models to the CRISM spectra. A good fit is obtained using the approximate size as ice crystals that have been detected by Phoenix using LIDAR (35 microns) and solar observations (10-70 microns) in the early morning period (Whiteway et al., 2009; Lemmon, 2014). It is conceivable that the snowflakes are deposited onto the cap in the early morning hours and are then observed by CRISM on its next observation of the cap. Our observations would then suggest that this process is more likely to occur as the atmosphere cools closer to the geophysical pole.

*5.3 Net Condensation and Net Sublimation Time Periods*

When we observe CRISM spectra from Point B as a function of time, we can identify two broad periods for the summer water cycle in the north polar region. The first part of summer is dominated by strong solar insolation, favoring net sublimation, while in the last part of summer the radiative cooling of the atmosphere favors net condensation of water vapor onto the cap.

Figure 8 shows a cartoon of the north polar water cycle in the two different summertime modes highlighted by this study. On the left is the *sublimation period* ($L_s$=90 to 125 for Point B), where the polar cap is in net sublimation mode, solar insolation is the dominant term and there is no condensation occurring on average. Condensation may occur, particularly at night, however the daily average will be net sublimation of water ice. Of course, small regions of net condensation may occur over the polar regions poleward of 87ºN where CRISM are extremely sparse.

Shown on the right of Figure 8 is the cap in *condensation period*, for Point B, this extends from $L_s$=125-167. Sublimation may take place, particularly during the middle of the day, but the daily average is net condensation. For colder locations, closer to the pole than Point B, or regions that experience significant periods in shadow, the condensation period will begin earlier in the summer. For regions on the periphery, the sublimation period can be expected to last longer.

*5.4 Martian Water Budget Implications*

*5.4.1 Martian Water budget in northern summer*

We can combine our estimates of the amount of water ice deposition with the estimates of water ice and vapor from previous studies to come up with a simplified water budget during Martian northern summer.

Smith (2004) used MGS TES data to produce global maps of atmospheric water vapor over several Martian years. Smith reported that a plume of water vapor of strength 40 pr-μm (e.g. his Figure 5) was present over the north polar cap from $L_s$=90-150, when it decreases markedly to around 20 pr-μm. In broad terms, this requires that roughly 20 pr-μm of water ice is either deposited on the polar cap or travels equatorward at $L_s$=150.

Bass et al. (2000) estimated 14 microns of water ice may have been deposited on the north polar cap (we disregard their coarse grain deposition estimate, which we believe to be an unlikely scenario after the results discussed herein). Using the Bass et al. and Smith estimates, this leaves 6 pr-μm to travel equatorward.

Our calculations suggest that throughout late summer, a net total of 0.64 x10$^9$ kg per day is deposited on the polar cap, including the amount deposited minus the amount sublimated. This is equivalent to a layer ~70 precipitable microns thick of water ice per day in the $L_s$=135-164 time period. This estimate is far larger than the amount of water ice removed from the atmosphere (~20 precipitable microns), and thus it highlights the importance of intra-cap water ice transport for explaining our deposition observations.

### 5.4.2 Comparison to south polar deposition

The scattering calculations presented in this study estimate that on average a layer of water ice ~70 microns thick is deposited in the $L_s$=135-164 period. This is far larger than the results of water ice deposition on the south pole during summer, where an average water ice layer 0.6-6 microns deep has been estimated using CRISM observations (Brown et al. (2014)). This result is undoubtedly due to the large amount of water vapor present over the north polar cap in comparison to the south polar cap (e.g. Smith (2004)).

### 5.4.3 Volatile measurements during polar night

How much water ice is deposited during the polar night? Since CRISM is a passive instrument, we have no data on water ice deposition during the polar night. We have to assume that further water ice deposition is likely after $L_s$=167, which is our last observation time for MY28. After this point, as the sun drops below the horizon in the Martian arctic, remote sensing of water ice deposition will require high resolution active sensors that can differentiate $CO_2$ and $H_2O$ ice. One such future instrument has recently been proposed (Brown et al., 2015) that will address this water ice deposition question, in addition to many other outstanding Martian climate and energy balance questions.

*5.5 GCM and Mesoscale modeling comparisons*

*5.5.1 Comparison to mesoscale models condensation and sublimation maps*

Numerical models of Martian water ice cycle generally take two depositional terms into account. The first is a snowfall term where ice forms in the atmosphere and falls directly on the surface. The second term in the models is direct deposition of water vapor from the atmosphere to the surface. As discussed previously, the CRISM measurements are sensitive to the sum of these two processes and cannot distinguish between them.

CRISM observations of $H_2O$ index and derived net sublimation and condensation regions over the $L_s$=132-167 period shown in Figure 7d can be compared (with some caveats) to Figure 12b of Tyler and Barnes (2014), which was generated by a simulation at a slightly earlier period ($L_s$=120). Unfortunately, insufficient CRISM observations of the $L_s$=120 period are available to construct a mosaic for this time period and mesoscale observations of the $L_s$=132-167 time period do not yet exist. However, to first order, the Tyler and Barnes simulation results are in accord with the water cycle model presented here. The Tyler and Barnes simulated sublimation regions are around the periphery of the cap and net condensation regions occur in the interior of the cap.

Two discrepancies between CRISM observations in Figure 7b and Figure 12 of Tyler and Barnes are important to note. Firstly, the north polar outlier deposits are undergoing net sublimation in the Tyler and Barnes model, whereas they are clearly undergoing net condensation in Figure 7d. This is largely due to the difference in time period, and could highlight another region of interest for future study. Secondly, the Tyler and Barnes model shows the strongest deposition occurring at $L_s$=120 over the direct pole, which is largely inaccessible to CRISM. The small number of cap overpasses shown in Figure 7a and 7b do not show unusually high amounts of deposition occurring poleward of 87°N during mid to late summer. It is difficult to draw strong conclusions from these handful of observations and more data is needed to settle the matter.

Quantitatively comparing the deposition estimates of Tyler and Barnes (2014) with our estimates, we see that their total deposition rates are ~0.5 x10$^9$ kg per day and sublimation rates of 8.9 x10$^9$ kg per day. Our deposition estimate is 0.11 x10$^9$ kg per day and our sublimation estimate is 0.05 x10$^9$ kg per day. Our best estimates of deposition rates in the net deposition region in late summer are ~140 microns per day. Tyler and Barnes report maximum deposition rates of 5 microns per day in mid summer. These differences are also largely due to the difference in season between the Tyler and Barnes simulation in mid summer ($L_s$=120) and our mapping season ($L_s$=135-164).

It is important to note that in addition to the offset in time period, the resolution of the mesoscale grid for Tyler and Barnes' study was 15km, compared with our roughly 0.5km per pixel resolution in the Figure 2-3 and 7. The results of this initial comparison certainly invite further mesoscale modeling during late summer in the north polar region.

*5.5.2 Direct deposition model and MAVAIL parameter*

As discussed in Tyler and Barnes (2014), the current state-of-the-art numerical models of the Martian surface-atmosphere transport employ a heuristic scheme to determine the amount of water ice deposited during each GCM time step. The flux of vapor at the surface ($F_{surfwater}$) is calculated to be the product of a mixing coefficient ($C_{mixing}$) and difference between the equilibrium mass mixing ratio over ice at ground temperature ($Mr_{equil}$) and the vapor mass mixing ratio at the lowest atmospheric layer ($Mr_{lowest}$). The sign of the difference term determines whether sublimation or condensation takes place.

$$F_{surfwater} = C_{mixing} \left( Mr_{equil} - Mr_{lowest} \right) \qquad (4)$$

The mixing ratio, $C_{mixing}$, is a function of air density, surface wind stress, a stability function particular to the GCM planetary boundary layer scheme and a moisture availability (*MAVAIL*) parameter. *MAVAIL* is a tunable parameter that reflects the 'ease' with which ice atoms stick to the Martian surface. Since it is not a physically measurable quantity, it is not possible to determine experimentally. Such a parameter is likely to be a function of ice and regolith porosity, chemical activation, grain shapes and sizes, the history of condensation and densification of the icepack (Arthern et al., 2000) and many other second order factors.

Tyler and Barnes (2014) chose values of *MAVAIL* = 1 for sublimation of ices from the north polar cap, 0.5 for sublimation from off-cap snow or frost, and following some experimentation, chose *MAVAIL*=0.1 for deposition onto any grid in the simulation. Tyler and Barnes only simulated a period of 20 sols around $L_s$=120 for their 2014 simulation, but in future numerical experiments, the tuning of the *MAVAIL* parameter could be informed by comparisons with the seasonal water deposition amounts reported in this paper.

## 6. Conclusions

This investigation has used CRISM mapping data to examine the summer water ice cycle of the Martian north polar cap. We have mapped changes in the $H_2O$ index and snow cap albedo as a function of space and time and identified net sublimation and condensation regions and periods at the limit of CRISM spatial and temporal resolution.

1. **Quantification of summer water ice deposition**. Under the assumption that the shrinking of apparent grain sizes is due to the deposition of fine grained water ice, we approximated the total amount of water ice deposited on the cap each summer. The scattering calculations presented herein suggest that on average a water ice layer ~70 microns thick is deposited in the late summer $L_s$=135-164 period. This is far larger than the results of deposition on the south pole during summer, where an average layer 0.6-6 microns deep has been estimated by Brown et al. (2014). This estimate is far higher than previous estimates by Bass et al. (2000) and highlights the far greater complexity of the north polar cap water cycle than can now be apprehended using the CRISM instrument multi-year dataset.

2. **Repeatable water cycle**. We have identified a repeatable summer cycle that displays two stable modes, 'net deposition' and 'net sublimation'. We have used individual CRISM spectra from small observations of the cap throughout the summer to show that the broad behavior of deposition and sublimation that we were able to map in MY28 extends to other Mars years and is therefore a regular feature of the current Martian climate.

3. **Mode flip timing**. We have established that regions of the cap will undergo a 'mode flip' depending (to first order) on their proximity to the north pole. Topography, composition and shadowing will also play an important local role.

4. **Net sublimation period**. In the interior of the Martian north polar cap, for example at Point B, water ice absorption bands grow from the end of springtime ($L_s$=85) until $L_s$=120. This will likely be due to a combination of 1.) thermal metamorphism of water ice (Eluszkiewicz, 1993; Ossipian and Brown, 2014) resulting in growth of grains as they age, and 2.) removal of a coating of fine grains by wind or sublimation, revealing larger grained ice beneath.

5. **Net deposition period**. In the cap interior at Point B, after $L_s$=135 (see Figure 4a and b) the absorption bands depths begin to decrease, consistent with the deposition of fine grained ice. This may occur through a combination of precipitation (snowfall) and/or direct deposition from water vapor in the atmosphere.

6. **Net deposition regions**. We have shown that the apparent patterns of water ice deposition on the north polar cap favors the interior of the polar cap and the Gemina Lingulae region of the cap. These areas enter the 'net deposition' phase earlier than regions on the periphery of the NPRC.

7. **Net sublimation regions**. We have observed that regions experiencing high insolation remain in net sublimation mode for more extended periods of time. For example, the periphery of the cap acts as a net sublimation region throughout summer.

# 7. Acknowledgements

We would like to thank two anonymous referees and Sylvain Piqueux for very helpful reviews that improved the paper considerably. We also thank Mike Wolff for providing his code and expertise on DISORT_multi which is used in the Appendix. We would also like to thank the entire CRISM Team, particularly the Science Operations team at JHU APL, and also the MARCI PI (Mike Malin) and his staff.

All CRISM data used in this paper is publicly available at the Planetary Data System (PDS) Geosciences node, at http://pds-geosciences.wustl.edu/.

This investigation was funded by NASA Grants NNX13AJ73G and NNX11AN41G in the Mars Data Analysis Program administered by Mitch Schulte.

# Appendix – Onset of Polar Hood Obscuration

In order to determine the north polar hood NIR opacities, which are measured by the CRISM instrument and different from visible opacities available from the MARCI dataset, we use the CRISM dataset itself to constrain the timing and thickness of the north polar hood. We do this in order to rule out atmospheric interference that would undermine our quantitative findings and to quantify the errors induced by atmospheric effects on the $H_2O$ index.

In MY28, at the start of the MRO CRISM mission, a large number of full resolution observations were obtained at the Phoenix landing site at 68.22N, 234.3E in preparation for the landing in summer of MY29. As shown in Table A.1, these run from $L_s$=162-181, and cover the period when the north polar hood develops. We therefore can use them to bootstrap a pole-wide background optical depth for atmospheric water ice.

Inspection of the CRISM observations in Table A.1 shows the amount of atmospheric water ice broadly increases as a function of time. The Phoenix observations were made in MY28, same year as the CRISM images that we used to create Figures 2, 3 and 7 and the fits shown in Figures 6a and b. We calculated the $H_2O$ ice index (which is a proxy for the 1.5 μm band), and these values are tabulated for each observation. There is no ground ice over the Phoenix site at this time, so the signature of water ice is only due to the atmospheric water ice of the north polar hood.

| FRT # | $L_s$ | $H_2O$ index | Qualitative Assessment | Derived NIR optical depth of $H_2O$ ice |
|-------|-------|--------------|------------------------|------------------------------------------|
| 3CAB | 162 | 0.011 | No water ice signature | ~0 |
| 3F9A | 171 | 0.038 | Patchy cloud apparent | 0.75 |
| 419C | 176 | 0.102 | 100% water ice cloud | 4.5 |
| 4395 | 181 | 0.087 | 100% water ice cloud | 3.5 |

Table A.1 - CRISM FRT observations overlapping the north polar cap over the Phoenix landing site in MY28, one Mars year before landing. Each spectrum was taken from the center of each image. The landing site is located at 68.2N, 234.3E, at a MOLA elevation of -4108.660 relative to the Martian aeroid.

In order to convert the $H_2O$ ice index into an optical thickness of atmospheric water ice, we used the DISORT_multi program which is used to realistically model the effect of water ice and dust in the Martian atmosphere on CRISM spectra (Wiseman et al., 2014; Wolff et al., 2009). We first extracted a surface albedo spectrum for the Phoenix landing site by running DISORT in inverse mode, and obtained a spectrum of a roughly "flat" albedo close to 0.25 and a 3 micron water band, ubiquitous in the spectrally featureless region of Acidalia Planitia (Horgan et al., 2009).

Using this "bland" spectrum as a surface albedo, and employing a Lambertian scattering assumption for the surface, we generated artificial Martian atmospheres using the DISORT_multi code for different water ice optical depths, at 1nm resolution. These are shown in Table A.2.

| Ice optical depth | $H_2O$ index |
|:---:|:---:|
| 0 | 0.008 |
| 0.05 | 0.017 |
| 0.1 | 0.018 |
| 0.2 | 0.021 |
| 0.25 | 0.023 |
| 0.5 | 0.027 |
| 0.75 | 0.036 |
| 1.0 | 0.041 |
| 2.0 | 0.063 |
| 3.0 | 0.079 |
| 4.0 | 0.095 |
| 4.5 | 0.103 |

Table A.2 – DISORT_multi results of varying water ice optical depth and resulting $H_2O$ index. The artificial atmospheres are generated over a featureless ground spectrum based on the Phoenix landing site, using solar incidence angle = 75°, emission angle=1°, phase angle =75°.

Using the data in Table A.2, we can fill out the final column in Table A.1, which assigns a derived water ice optical depth for the near infrared part of the spectrum, where the 1.5 µm band is located and which we have relied upon in our quantitative fitting in this paper. At $L_s$=162, the atmosphere has NIR optical opacity of approximately 0. At $L_s$=171, this increases markedly to around 0.75, then rapidly climbs to 4.5 at $L_s$=176, then back to 3.5 for $L_s$=181. The peak at $L_s$=176 is most likely due to baroclinic cloud activity.

This method makes the assumption that the north polar hood is relatively homogenous and does not increase in optical thickness from the Phoenix landing site at 68°N to the polar cap at Point B at 85°N. MARCI images support this assertion with strong concentrations of baroclinic waves circling around the edges of the pole, and Figure 3c of Benson et al. (2011) in fact indicates that the hood is actually thicker at the edges and less optically thick at the north pole, making our assumptions here appear conservative.

Based on a linear extrapolation between the derived optical depths shown in the final column of Table A.1 (assuming a steady build up of cloud optical depth from $L_s$=162-171) and considering a NIR optical depth of greater than 0.25 to be 'obscuring', we find that the north polar hood is obscuring in the NIR beyond $L_s$=165.

Given that the $H_2O$ ice index of an almost perfectly clear observation at $L_s$=162 (first row) was still 0.011, we consider this a good quantification of the error induced in the $H_2O$ index by the atmosphere. Therefore, we adopt at ± 0.01 error for the $H_2O$ index due to atmospheric effects.

# REFERENCES

Arthern, R.J., Winebrenner, D.P., Waddington, E.D., 2000. Densification of Water Ice Deposits on the Residual North Polar Cap of Mars. Icarus 144, 367–381.

Bass, D.S., Herkenhoff, K.E., Paige, D.A., 2000. Variability of Mars' North Polar Water Ice Cap: I. Analysis of Mariner 9 and Viking Orbiter Imaging Data. Icarus 144, 382–396.

Bass, D.S., Paige, D.A., 2000. Variability of Mars' North Polar Water Ice Cap: II. Analysis of Viking IRTM and MAWD Data. Icarus 144, 397–409.

Becerra, P., Byrne, S., Brown, A.J., 2015. Transient Bright "Halos" on the South Polar Residual Cap of Mars: Implications for Mass-Balance. Icarus 211–225. doi:10.1016/j.icarus.2014.04.050

Benson, J.L., Kass, D.M., Kleinbˆhl, A., 2011. Mars' north polar hood as observed by the Mars Climate Sounder. J. Geophys. Res. 116, E03008.

Bibring, J.-P., Langevin, Y., Poulet, F., Gendrin, A., Gondet, B., Berthe, M., Soufflot, A., Drossart, P., Combes, M., Bellucci, G., Moroz, V., Mangold, N., Schmitt, B., 2004. Perennial water ice identified in the south polar cap of Mars. Nature 428, 627–630.

Brown, A.J., 2006. Spectral Curve Fitting for Automatic Hyperspectral Data Analysis. IEEE Trans. Geosci. Remote Sens. 44, 1601–1608. doi:10.1109/TGRS.2006.870435

Brown, A.J., Byrne, S., Tornabene, L.L., Roush, T., 2008a. Louth Crater: Evolution of a layered water ice mound. Icarus 196, 433–445.

Brown, A.J., Calvin, W.M., McGuire, P.C., Murchie, S.L., 2010. Compact Reconnaissance Imaging Spectrometer for Mars (CRISM) south polar mapping: First Mars year of observations. J. Geophys. Res. 115, doi:10.1029/2009JE003333.

Brown, A.J., Calvin, W.M., Murchie, S.L., 2012. Compact Reconnaissance Imaging Spectrometer for Mars (CRISM) north polar springtime recession mapping: First 3 Mars years of observations. J. Geophys. Res. 117, E00J20.

Brown, A.J., Michaels, T.I., Byrne, S., Sun, W., Titus, T.N., Colaprete, A., Wolff, M.J., Videen, G., Grund, C.J., 2015. The case for a modern multiwavelength, polarization-sensitive LIDAR in orbit around Mars. J. Quant. Spectrosc. Radiat. Transf. 153, 131–143. doi:10.1016/j.jqsrt.2014.10.021

Brown, A.J., Piqueux, S., Titus, T.N., 2014. Interannual observations and quantification of summertime H2O ice deposition on the Martian CO2 ice south polar cap. Earth Planet. Sci. Lett. 406, 102–109. doi:10.1016/j.epsl.2014.08.039

Brown, A.J., Sutter, B., Dunagan, S., 2008b. The MARTE Imaging Spectrometer Experiment: Design and Analysis. Astrobiology 8, 1001–1011. doi:10.1089/ast.2007.0142

Byrne, S., 2009. The Polar Deposits of Mars. Annu. Rev. Earth Planet. Sci. 37, 535–560.

Byrne, S., Zuber, M.T., Neumann, G.A., 2008. Interannual and seasonal behavior of Martian Residual Ice-Cap Albedo. Planet. Space Sci. 56, 194–211.

Calvin, W.M., James, P.B., Cantor, B.A., Dixon, E.M., 2014. Interannual and seasonal changes in the north polar ice deposits of Mars: Observations from MY 29–31 using MARCI. Icarus.

Calvin, W.M., Titus, T.N., 2008. Summer season variability of the north residual cap of Mars as observed by the Mars Global Surveyor thermal emission spectrometer (MGS-TES). Planet. Space Sci. 56, 212–226.

Cantor, B.A., James, P.B., Calvin, W.M., 2011. MARCI and MOC observations of the atmosphere and surface cap in the north polar region of Mars. Icarus 208, 61–81.

Cantor, B., Malin, M., Edgett, K.S., 2002. Multiyear Mars Orbiter Camera (MOC) observations of repeated Martian weather phenomena during the northern summer season. J. Geophys. Res.-Planets 107, art. no.–5014.

Carmagnola, C.M., Domine, F., Dumont, M., Wright, P., Strellis, B., Bergin, M., Dibb, J., Picard, G., Libois, Q., Arnaud, L., Morin, S., 2013. Snow spectral albedo at Summit, Greenland: measurements and numerical simulations based on physical and chemical properties of the snowpack. The Cryosphere 7, 1139–1160.

Clancy, R.T., Lee, S.W., 1991. A new look at dust and clouds in the Mars atmosphere - Analysis of emission-phase-function sequences from global Viking IRTM observations. Icarus 93, 135–158.

Clancy, R.T., Sandor, B.J., Wolff, M.J., Christensen, P.R., Smith, M.D., Pearl, J.C., Conrath, B.J., Wilson, R.J., 2000. An intercomparison of ground-based millimeter, MGS TES, and Viking atmospheric temperature measurements: Seasonal and interannual variability of temperatures and dust loading in the global Mars atmosphere. J. Geophys. Res.-Planets 105, 9553–9571.

Clifford, S.M., 1993. A Model for the Hydrologic and Climatic Behavior of Water on Mars. J. Geophys. Res.-Planets 98, 10973–11016.

Eluszkiewicz, J., 1993. On the Microphysical State of the Martian Seasonal Caps. Icarus 103, 43–48.

Farmer, C.B., Davies, D.W., Holland, A.L., D.D., L., Doms, P.E., 1977. Mars: Water vapour observations from the Viking Orbiters. J. Geophys. Res. 82, 4225–4248.

Farmer, C.B., Davies, D.W., Laporte, D.D., 1976. Mars: Northern Summer Ice Cap—Water Vapor Observations from Viking 2. Science 194, 1339–1341.

Fishbaugh, K.E., Head, J.W., 2005. Origin and characteristics of the Mars north polar basal unit and implications for polar geologic history. Icarus 174, 444–474.

Fouchet, T., Lellouch, E., Ignatiev, N.I., Forget, F., Titov, D.V., Tschimmel, M., Montmessin, F., Formisano, V., Giuranna, M., Maturilli, A., others, 2007. Martian water vapor: Mars Express PFS/LW observations. Icarus 190, 32–49.

French, R.G., Gierasch, P.J., 1979. The Martian Polar Vortex: Theory of Seasonal Variation and Observations of Eolian Features. J. Geophys. Res. 84, 4634–4642.

Grenfell, T.C., 1991. A radiative transfer model for sea ice with vertical structure variations. J. Geophys. Res. Oceans 96, 16991–17001. doi:10.1029/91JC01595

Grenfell, T.C., Perovich, D.K., Ogren, J.A., 1981. Spectral albedos of an alpine snowpack. Cold Reg. Sci. Technol. 4, 121–127.

Grenfell, T.C., Warren, S.G., Mullen, P.C., 1994. Reflection of solar radiation by the Antarctic snow surface at ultraviolet, visible, and near-infrared wavelengths. J. Geophys. Res. 99, 18669–18684.

Haberle, R.M., Jakosky, B.M., 1990. Sublimation and transport of water from the north residual polar cap on Mars. J. Geophys. Res. 95, 1423–1437.

Hamre, B., Winther, J.-G., Gerland, S., Stamnes, J.J., Stamnes, K., 2004. Modeled and measured optical transmittance of snow-covered first-year sea ice in Kongsfjorden, Svalbard. J. Geophys. Res. Oceans 109, C10006. doi:10.1029/2003JC001926

Hart, H.M., Jakosky, B.M., 1986. Composition and stability of the condensate observed at the Viking Lander 2 site on Mars. Icarus 66, 134–142.

Horgan, B.H., Bell, J.F., III, Noe Dobrea, E.Z., Cloutis, E.A., Bailey, D.T., Craig, M.A., Roach, L.H., Mustard, J.F., 2009. Distribution of hydrated minerals in the north polar region of Mars. J. Geophys. Res. 114.

Houben, H., R. M. Haberle, R. E. Young, and A. P. Zent. "Evolution of the Martian Water Cycle." Advances in Space Research 19 (1997): 1233–36.

Jin, Z., Stamnes, K., Weeks, W.F., Tsay, S.-C., 1994. The effect of sea ice on the solar energy budget in the atmosphere-sea ice-ocean system: A model study. J. Geophys. Res. Oceans 1978–2012 99, 25281–25294.

Kahn, R., 1984. The spatial and seasonal distribution of Martian clouds and some meteorological implications. J. Geophys. Res. 89, 6671–6688.

Kaufmann, E., Hagermann, A., 2015. Penetration of solar radiation into pure and Mars-dust contaminated snow. Icarus 252, 144–149. doi:10.1016/j.icarus.2015.01.007

Kieffer, H., 1990. H2O grain size and the amount of dust in Mars' residual north polar cap. J. Geophys. Res. 95, 1481–1493.

Kieffer, H., 1987. How dirty is Mars' north polar cap, and why isn't it black? Presented at the MEVTV Workshop on Nature and Composition of Surface Units on Mars, LPI, Houston, TX, pp. 72–73.

Kieffer, H.H., 2007. Cold jets in the Martian polar caps. J. Geophys. Res. 112, 10.1029/2006JE002816.

Kieffer, H.H., Chase, S.C., Martin, T.Z., Miner, E.D., Palluconi, F.D., 1976. Martian North Pole Summer Temperatures: Dirty Water Ice. Science 194, 1341–1344.

Kieffer, H.H., Titus, T.N., 2001. TES Mapping of Mars' North Seasonal Cap. Icarus 154, 162–180.

Langevin, Y., Bibring, J.-P., Montmessin, F., Forget, F., Vincendon, M., Douté, S., Poulet, F., Gondet, B., 2007. Observations of the south seasonal cap of Mars during recession in 2004–2006 by the OMEGA visible/near-infrared imaging spectrometer on board Mars Express. J. Geophys. Res. 112, 10.1029/2006JE002841.

Langevin, Y., Poulet, F., Bibring, J.-P., Schmitt, B., Doute, S., Gondet, B., 2005. Summer Evolution of the North Polar Cap of Mars as Observed by OMEGA/Mars Express. Science 307, 1581–1584.

Lemmon, M., 2014. Large water ice aerosols in martian north polar clouds. Presented at the Fifth International Conference on the Mars Atmosphere: Modeling and Observations, LPI, Oxford, England.

Lemmon, M.T., 2014. Large water ice aerosols in martian north polar clouds. Presented at the Fifth International Conference on the Mars Atmosphere: Modeling and Observations, Oxford, England.

Lian, Y., Richardson, M.I., Newman, C.E., Lee, C., Toigo, A.D., Mischna, M.A., Campin, J.-M., 2012. The Ashima/MIT Mars GCM and argon in the martian atmosphere. Icarus 218, 1043–1070.

Malin, M.C., Bell, J.F., Calvin, W., Clancy, R.T., Haberle, R.M., James, P.B., Lee, S.W., Thomas, P.C., Caplinger, M.A., 2001. Mars Color Imager (MARCI) on the Mars Climate Orbiter. J. Geophys. Res.-Planets 106, 17651–17672.

Murchie, S., Arvidson, R., Bedini, P., Beisser, K., Bibring, J.-P., Bishop, J., Boldt, J., Cavender, P., Choo, T., Clancy, R.T., Darlington, E.H., Des Marais, D., Espiritu, R., Fort, D., Green, R., Guinness, E., Hayes, J., Hash, C., Heffernan, K., Hemmler, J., Heyler, G., Humm, D., Hutcheson, J., Izenberg, N., Lee, R., Lees, J., Lohr, D., Malaret, E., T., M., McGovern, J.A., McGuire, P., Morris, R., Mustard, J., Pelkey, S., Rhodes, E., Robinson, M., Roush, T., Schaefer, E., Seagrave, G., Seelos, F., Silverglate, P., Slavney, S., Smith, M., Shyong, W.-J., Strohbehn, K., Taylor, H., Thompson, P., Tossman, B., Wirzburger, M., Wolff, M., 2007. Compact Reconnaissance Imaging Spectrometer for Mars (CRISM) on Mars Reconnaissance Orbiter (MRO). J. Geophys. Res. 112, E05S03, doi:10.1029/2006JE002682.

Nelder, J.A., Mead, R., 1965. A simplex method for function minimization. Comput. J. 7, 308–313.

Ossipian, S., Brown, A., 2014. How fast can water ice grains grow on the summertime Martian North Pole? Presented at the Eigth International Conference on Mars, Caltech, Pasadena, CA, p. Abs. #1071.

Pankine, A.A., Tamppari, L.K., Smith, M.D., 2010. MGS TES observations of the water vapor above the seasonal and perennial ice caps during northern spring and summer. Icarus 210, 58–71.

Pilorget, C., Forget, F., Millour, E., Vincendon, M., Madeleine, J.B., 2011. Dark spots and cold jets in the polar regions of Mars: new clues from a thermal model of surface $CO_2$ ice. Icarus 213, 131–149.

Pollack, J.B., Haberle, R.M., 1990. Simulations of the General Circulation of the Martian Atmosphere: 1. Polar Processes. J. Geophys. Res. 95, 1447–1473.

Putzig, N.E., Phillips, R.J., Campbell, B.A., Holt, J.W., Plaut, J.J., Carter, L.M., Egan, A.F., Bernardini, F., Safaeinili, A., Seu, R., 2009. Subsurface structure of Planum Boreum from Mars Reconnaissance Orbiter Shallow Radar soundings. Icarus 204, 443–457.

Reutlinger, A., Schurz, G., Hüttemann, A., 2011. Ceteris Paribus Laws.

Roush, T., Pollack, J.B., Orenberg, J., 1991. Derivation of Midinfrared (5-25 micrometer) Optical Constants of Some Sillicates and Palagonite. Icarus 94, 191–208.

Smith, M.D., 2004. Interannual variability in TES atmospheric observations of Mars during 1999-2003. Icarus 167, 148–165.

Smith, M.D., Wolff, M.J., Clancy, R.T., Kleinböhl, A., Murchie, S.L., 2013. Vertical distribution of dust and water ice aerosols from CRISM limb-geometry observations. J. Geophys. Res. Planets 118, 321–334.

Stamnes, K., Hamre, B., Stamnes, J.J., Ryzhikov, G., Biryulina, M., Mahoney, R., Hauss, B., Sei, A., 2011. Modeling of radiation transport in coupled atmosphere-snow-ice-ocean systems. J. Quant. Spectrosc. Radiat. Transf. 112, 714–726.

Stamnes, K., Tsay, S.C., Wiscombe, W.J., Jayaweera, K., 1988. Numerically stable algorithm for discrete-ordinate-method radiative transfer in multiple scattering and emitting layered media. Appl. Opt. 27, 2502–2509.

Tamppari, L.K., Smith, M.D., Bass, D.S., Hale, A.S., 2007. Water-ice clouds and dust in the north polar region of Mars using MGS TES data. Planet. Space Sci. 56, 227–245. doi:doi:10.1016/j.pss.2007.08.011

Titus, T.N., W.M. Calvin, H.H. Kieffer, Y. Langevin, and T.H. Prettyman. "Martian Polar Processes." In The Martian Surface: Composition, Mineralogy, and Physical Properties, edited by J.F. Bell, 578–98. Cambridge University Press, 2008.

Tyler, D., Barnes, J.R., 2005. A mesoscale model study of summertime atmospheric circulations in the north polar region of Mars. J. Geophys. Res. Planets 110, E06007.

Tyler Jr., D., Barnes, J.R., 2014. Atmospheric mesoscale modeling of water and clouds during northern summer on Mars. Icarus 237, 388–414. doi:10.1016/j.icarus.2014.04.020

Wang, H., Ingersoll, A.P., 2002. Martian clouds observed by Mars Global Surveyor Mars Orbiter Camera. J. Geophys. Res. Planets 107, 5078. doi:10.1029/2001JE001815

Warren, S.G., 2013. Can black carbon in snow be detected by remote sensing? J. Geophys. Res. Atmospheres 118, 779–786. doi:10.1029/2012JD018476

Warren, S.G., 1984. Optical constants of ice from the ultraviolet to the microwave. Appl. Opt. 23, 1206–1225.

Warren, S.G., 1982. Optical properties of snow. Rev. Geophys. 20, 67–89.

Warren, S.G., Wiscombe, W.J., 1980. A Model for the spectral albedo of snow: II: snow containing atmospheric aerosols. J. Atmospheric Sci. 37, 2734–2745.

Whiteway, J.A., Komguem, L., Dickinson, C., Cook, C., Illnicki, M., Seabrook, J., Popovici, V., Duck, T.J., Davy, R., Taylor, P.A., Pathak, J., Fisher, D., Carswell, A.I., Daly, M., Hipkin, V., Zent, A.P., Hecht, M.H., Wood, S.E., Tamppari, L.K., Renno, N., Moores, J.E., Lemmon, M.T., Daerden, F., Smith, P.H., 2009. Mars Water-Ice Clouds and Precipitation. Science 325, 68–70. doi:10.1126/science.1172344

Wiseman, S.M., Arvidson, R.E., Wolff, M.J., Smith, M.D., Seelos, F.P., Morgan, F., Murchie, S.L., Mustard, J.F., Morris, R.V., Humm, D., McGuire, P.C., 2014. Characterization of Artifacts Introduced by the Empirical Volcano-Scan Atmospheric Correction Commonly Applied to CRISM and OMEGA Near-Infrared Spectra. Icarus. doi:10.1016/j.icarus.2014.10.012

Wolff, M.J., Smith, M.D., Clancy, R.T., Arvidson, R.E., Kahre, M., Seelos, F., Murchie, S., Savijarvi, H., 2009. Wavelength Dependence of Dust Aerosol Single Scattering Albedo As Observed by CRISM. J. Geophys. Res. 114, doi:10.1029/2009JE003350.

# FIGURES AND CAPTIONS

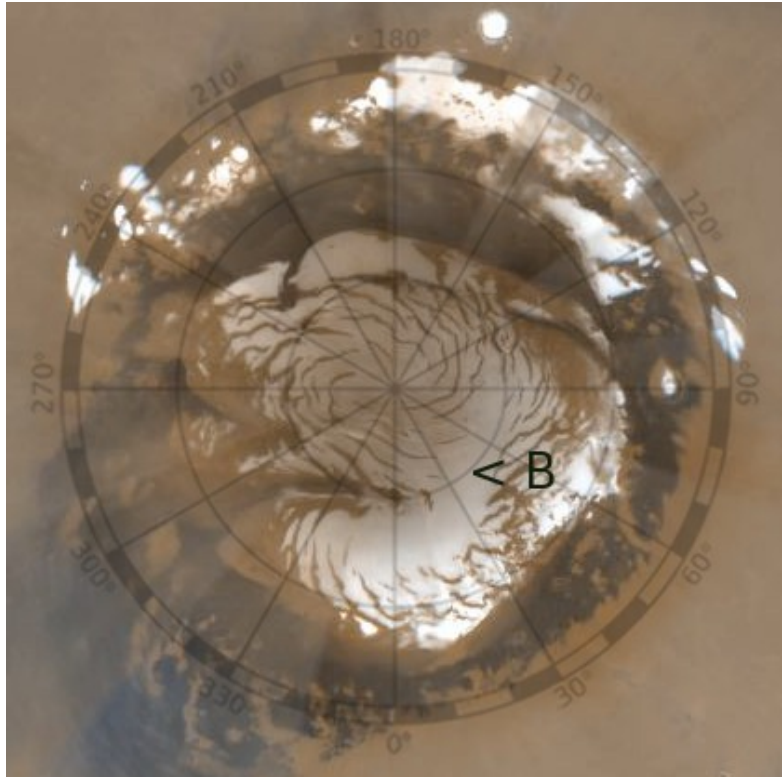

Figure 1 – MARCI image of north polar cap in MY 28 at $L_s$=130 (mid summer) showing location of CRISM spectra from Figure 4 (Point B). Outermost latitude circle is 75°N.

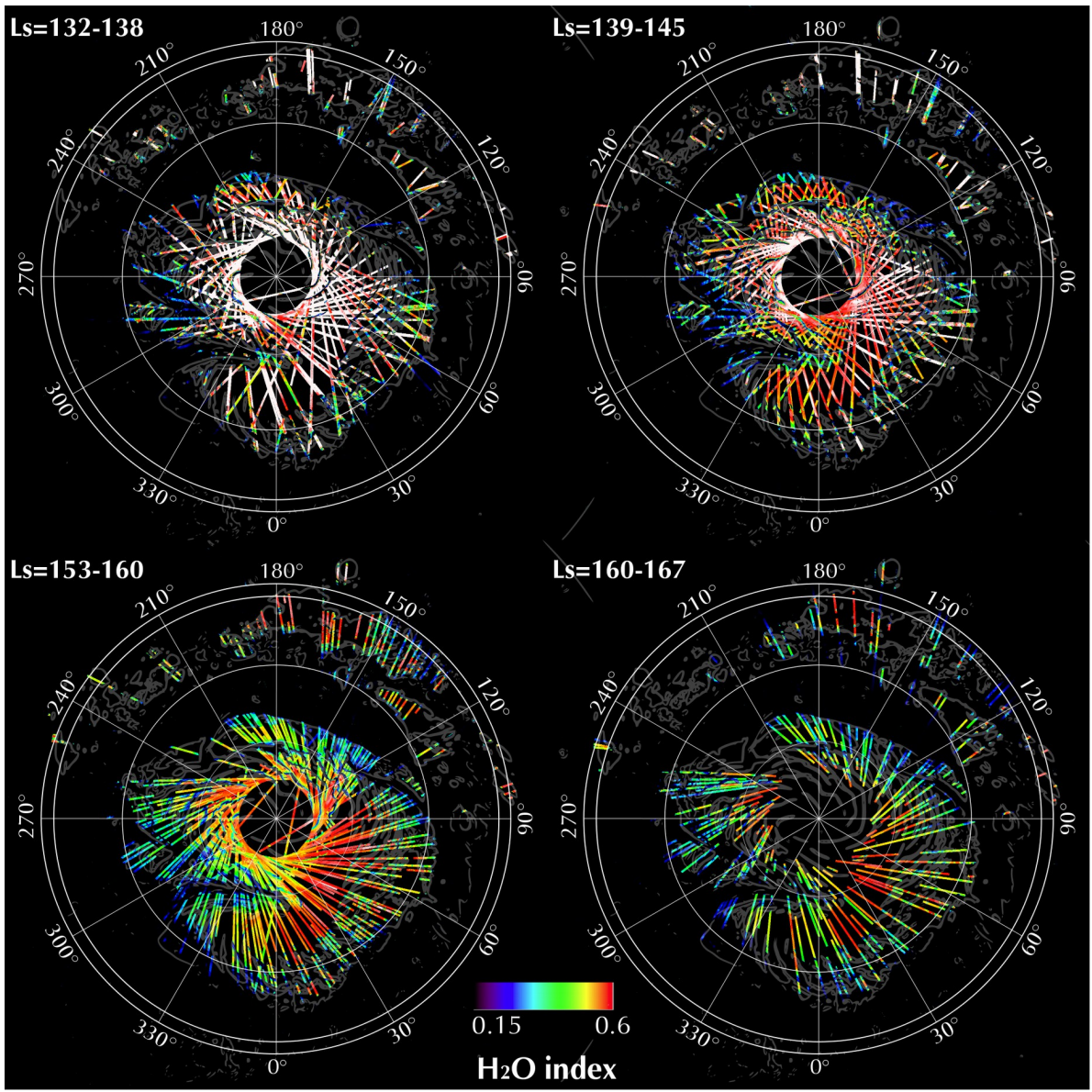

Figure 2 – MY 28 northern summer H₂O ice index mosaics. Note high H₂O index over the polar ice cap that decreases as summer progresses. Outermost latitude circle is 75°N. See text for discussion.

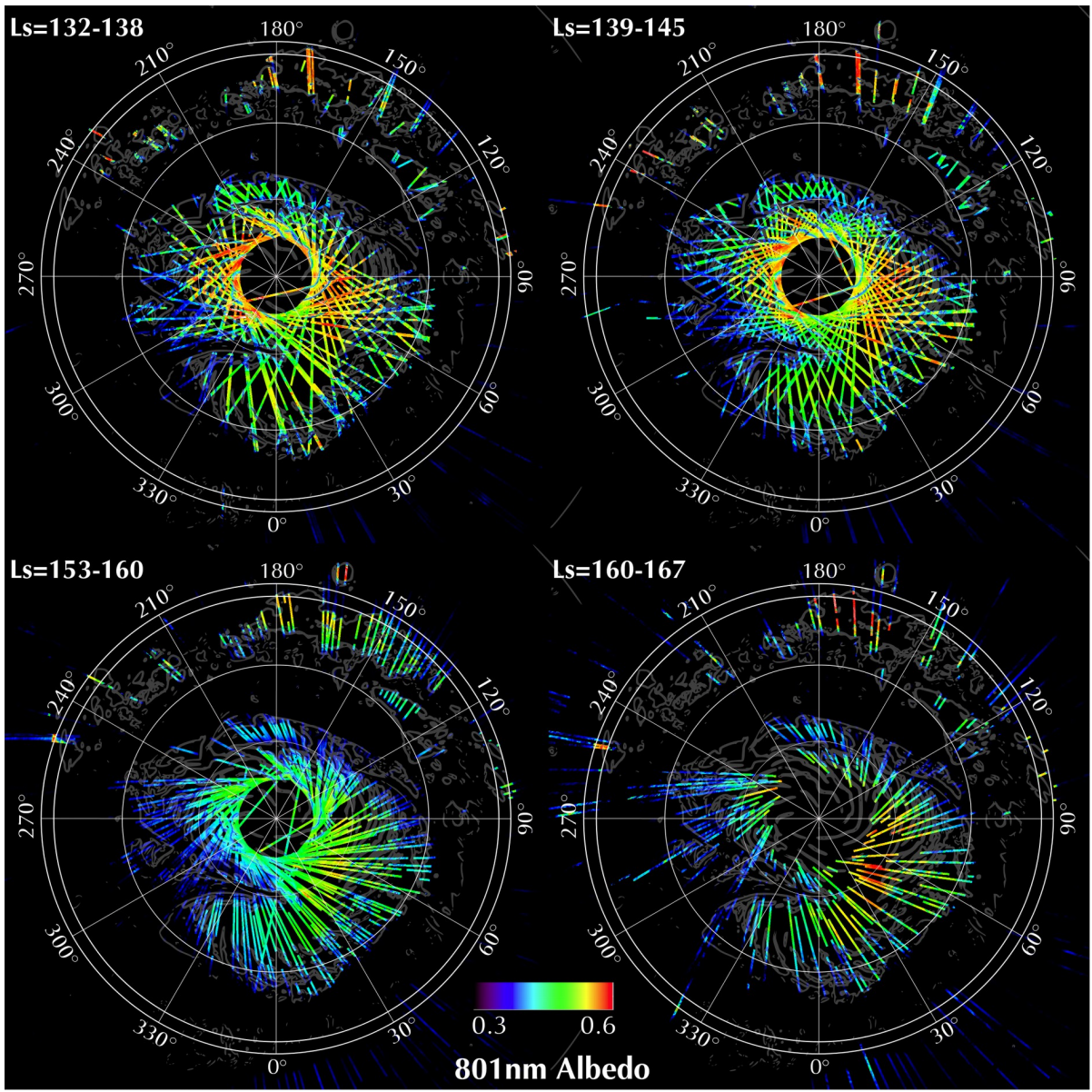

Figure 3 – MY 28 northern summer 801nm albedo mosaics. Note albedo is relatively steady (compared to the $H_2O$ index in Fig. 2) over the polar ice cap and slightly increases in the last panel. Outermost latitude circle is 75°N. See text for discussion.

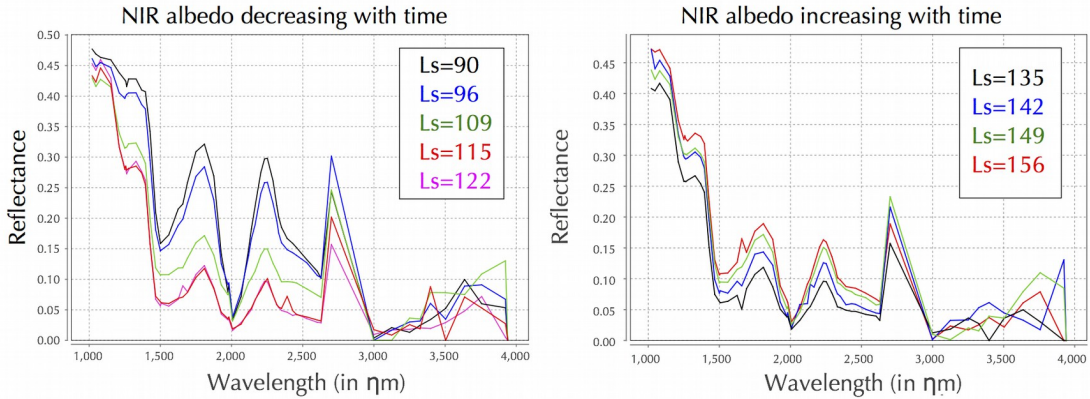

Figure 4 – CRISM spectra from Point B for summer of MY 29. All spectra are from x=550,y=550 in our 1000x1000 pixel mosaics. Spectra on left show decreasing albedo (increasing $H_2O$ index) and those on the right show increasing albedo (decreasing $H_2O$ index) with time. Note lowest 1.5 µm albedo (largest $H_2O$ index) is achieved both years in $L_s$=122-135 period.

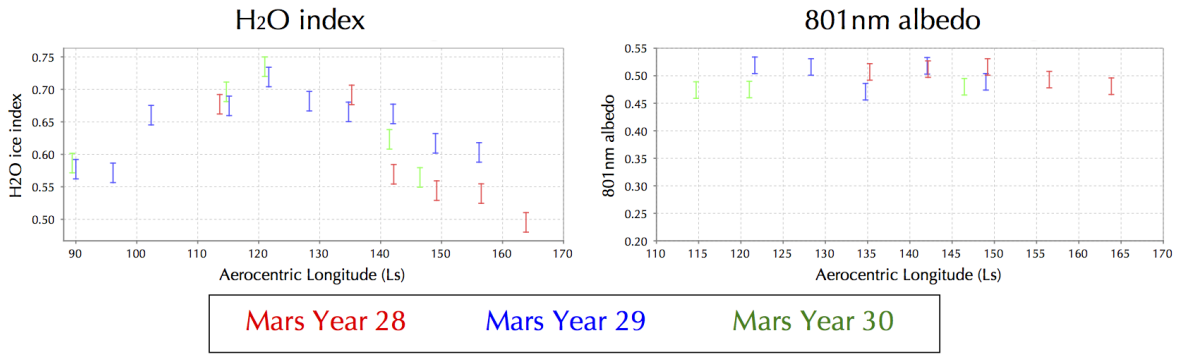

Figure 5 – (left) CRISM Mapping spectra $H_2O$ ice index taken from points close to Point B from MY 28-30 during $L_s$=90-170 (early-late summer). Note decrease in $H_2O$ index after $L_s$=120 is apparent across all three Mars years. (right) CRISM

spectra 801nm albedo (which is relatively constant) for $L_s$=110-170 for same locations and Mars years.

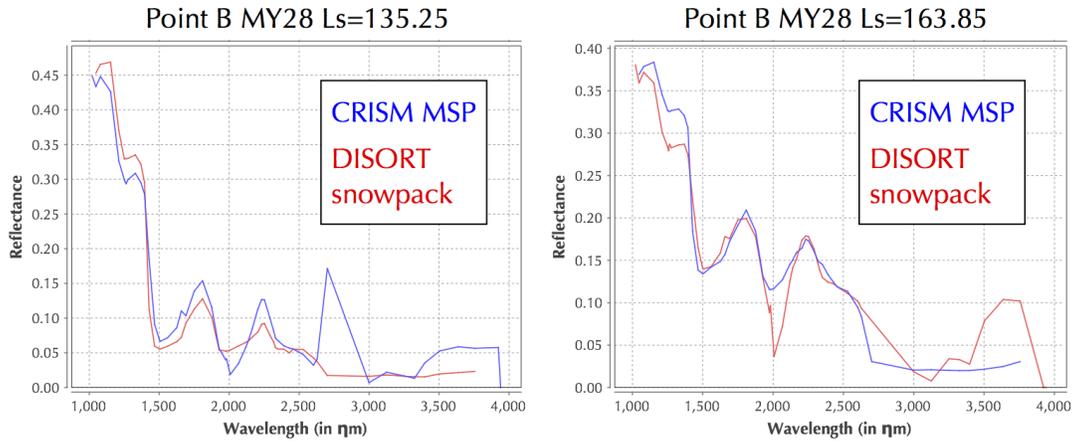

Figure 6a – (left) Point B spectrum from MY 28 $L_s$=135.25 (blue) compared with DISORT snowpack models of dust mixed with water ice (red). Figure 6b (right) As for Fig. 6a, however the season is late summer ($L_s$=163.85) and the fit includes a layer of fine grained ice resulting in a higher infrared albedo. An atmospheric $CO_2$ band at 2000 ηm is responsible for the poor fit in that region.

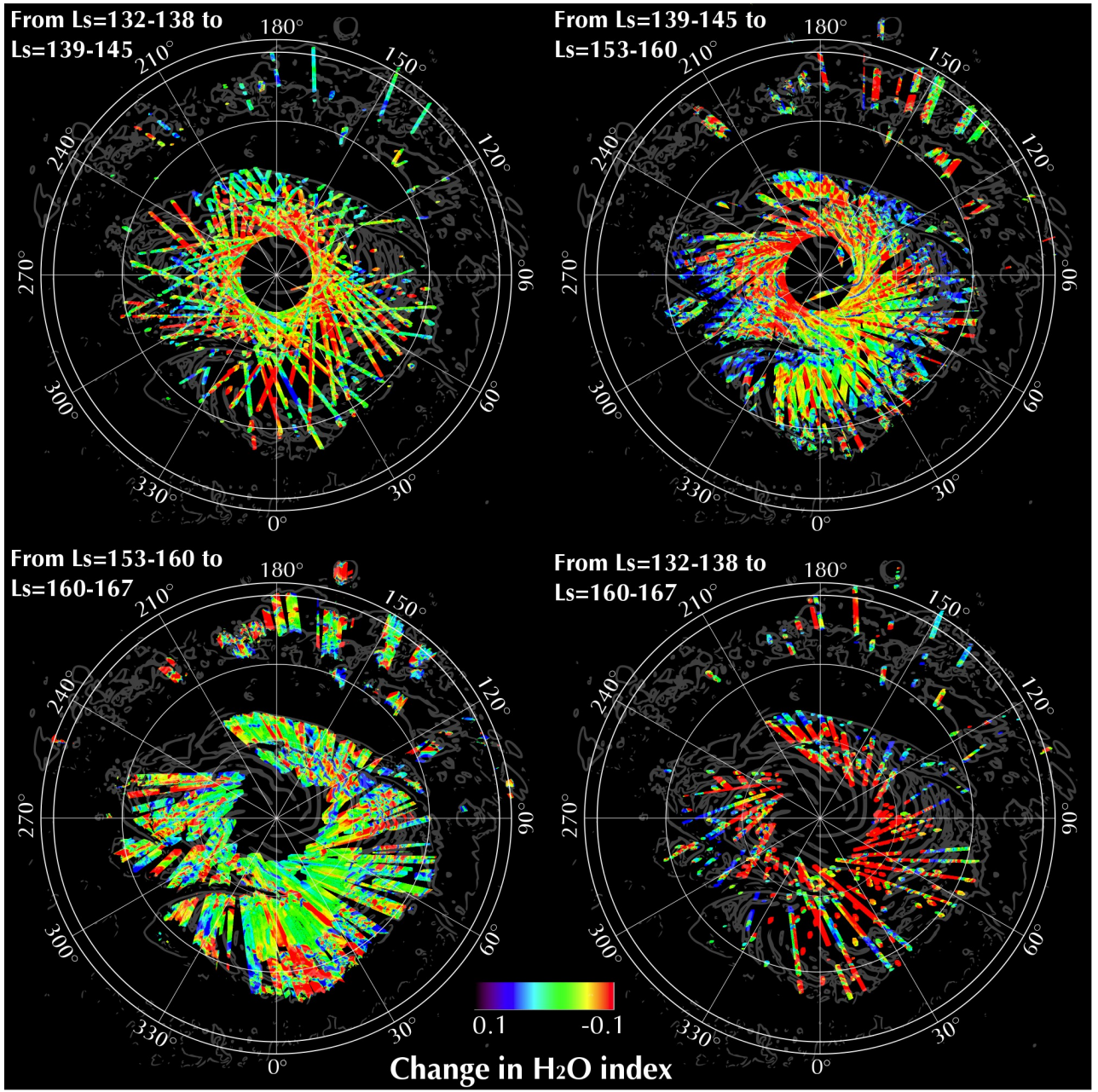

Figure 7 – Difference maps showing the change in $H_2O$ index between each mosaic in Figure 2. The bottom right image shows the $H_2O$ index difference across the entire set of mosaics. Red indicates a decrease in $H_2O$ index, blue indicates an increase in $H_2O$ index. For these difference maps, an interpolation method was used in order to increase the coverage of each CRISM swath. See text for further discussion.

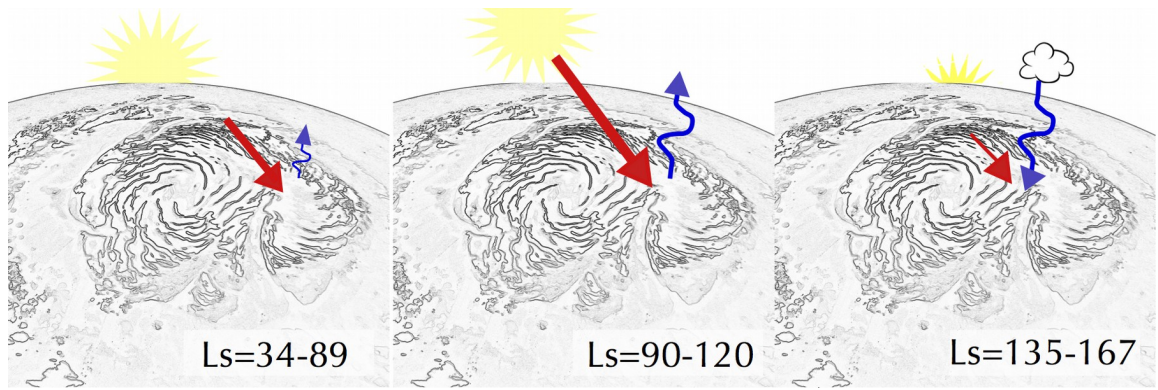

Figure 8 - A diagrammatic representation of key elements of the north polar water cycle as revealed by this study. On the left is the period after $CO_2$ ice has disappeared in mid spring. During this period the water ice band depths are stable, with only a small amount of sublimation (see Figure 4 of Brown et al,. 2012). In the center is the 'net sublimation' period, indicated by the wavy blue up arrow. On the right is 'net condensation' periods, shown by a wavy blue down arrow with a cloud that indicates snowfall (and direct deposition) is occurring during this time. The timings shown are appropriate for Point B, which is pointed to by the three red arrows. The length of the red arrows represents the strength of the solar insolation during each period. The timing of the 'mode flip' occurs earlier for regions closer to the pole than Point B, and later (or not at all) for regions on the edge of the cap.